# PET/CT Radiomic Sequencer for Prediction of EGFR and KRAS Mutation Status in NSCLC Patients

Isaac Shiri[1,2], Hassan Maleki[1,3], Ghasem Hajianfar[1], Hamid Abdollahi[1,4], Saeed Ashrafinia[5,6], Mostafa Ghelich Oghli[1], Mathieu Hatt[7], Mehrdad Oveisi[1,8], Arman Rahmim*[5,9,10]

1. Department of Biomedical and Health Informatics, Rajaie Cardiovascular Medical and Research Center, Iran University of Medical Science, Tehran, Iran
2. Research Center for Molecular and Cellular Imaging, Tehran University of Medical Sciences, Tehran, Iran
3. Department of Computer Science and Engineering, Shahid Beheshti University, Tehran, Iran
4. Department of Radiologic Sciences and Medical Physics, Faculty of Allied Medicine, Kerman University
5. Department of Radiology and Radiological Science, Johns Hopkins University, Baltimore MD, USA
6. Department of Electrical and Computer Engineering, Johns Hopkins University, Baltimore MD, USA
7. INSERM, UMR 1101, LaTIM, Univ Brest, F-29238, Brest, France
8. Department of Computer Science, University of British Columbia, Vancouver BC, Canada
9. Departments of Radiology and Physics & Astronomy, University of British Columbia, Vancouver BC, Canada
10. Department of Integrative Oncology, BC Cancer Research Centre, Vancouver BC, Canada

**First Author:**

Isaac Shiri is with the Biomedical and Health Informatics Department, Rajaie Cardiovascular Medical and Research Center, Tehran, Iran (e-mail: Isaac.sh92@gmail.com)

**Corresponding Author:**

Arman Rahmim is with the Radiology and Physics & Astronomy Departments, University of British Columbia, Vancouver, Canada. (e-mail: arman.rahmim@ubc.ca)

Associate Professor of Radiology and Physics, University of British Columbia

Senior Scientist & Provincial Medical Imaging Physicist, BC Cancer Agency

BC Cancer Research Center

675 West 10th Ave

Office 5-114

Vancouver, BC, V5Z 1L3

http://rahmimlab.com

Phone: 604-675-8262

Email: arman.rahmim@ubc.ca

**Abstract:**

The aim of this study was to develop radiomic models using PET/CT radiomic features with different machine learning approaches for finding best predictive epidermal growth factor receptor (EGFR) and Kirsten rat sarcoma viral oncogene (KRAS) mutation status. Patient's images including PET and CT [diagnostic (CTD) and low dose CT (CTA)] were pre-processed using wavelet (WAV), Laplacian of Gaussian (LOG) and 64 bin discretization (BIN) (alone or in combinations) and several features from images were extracted. The prediction performance of model was checked using the area under the receiver operator characteristic (ROC) curve (AUC). Results showed a wide range of radiomic model AUC performances up to 0.75 in prediction of EGFR and KRAS mutation status. Combination of K-Best and variance threshold feature selector with logistic regression (LREG) classifier in diagnostic CT scan led to the best performance in EGFR (CTD-BIN+B-KB+LREG, AUC: 0.75±0.10) and KRAS (CTD-BIN-LOG-WAV+B-VT+LREG, AUC: 0.75±0.07) respectively. Additionally, incorporating PET, kept AUC values at ~0.74. When considering conventional features only, highest predictive performance was achieved by PET SUVpeak (AUC: 0.69) for EGFR and by PET MTV (AUC: 0.55) for KRAS. In comparison with conventional PET parameters such as standard uptake value, radiomic models were found as more predictive. Our findings demonstrated that non-invasive and reliable radiomics analysis can be successfully used to predict EGFR and KRAS mutation status in NSCLC patients.

## I. INTRODUCTION

MOLECULAR Profiling is a standard protocol for better management of non-small cell lung cancer (NSCLC) patients. Analyses of the mutation status of epidermal growth factor receptor (EGFR) and Kirsten rat sarcoma viral oncogene (KRAS) mutations are frequently performed and used as management tools in non-small cell lung cancer (NSCLC) [1]. Accumulating evidence has suggested that mutations in KRAS and EGFR are considered as first lines for clinical decision making in NSCLS treatment and outcome improvement [2].

Radiomics is a new advanced approach aiming to find correlation between features extracted from medical images and clinical/biological data[3]. Image features have the potential to be used in a wide range of clinical applications including tumor characterization, staging, grading, therapy response assessment and prediction[4]. Based on advanced radiomics studies, converting images to high-dimensional, mineable, and quantitative data could provide an advanced, low cost and noninvasive to improve decision-support in clinics [5]. However radiomics studies suffer from several challenges, and a robust framework for clinical decision making is highly desired [6-11].

Studies on radiomic modelling have been made by using advanced algorithms such as machine learning (ML) approaches [12]. Previous works have tested several ML algorithms and identified that some algorithms can contribute to build highly accurate and reliable predictive and prognostic radiomic models [12].

Recently some studies have indicated that imaging features extracted from CT scan images could predict mutation status in NSCLC patients. Velazquez *et al*. [13] developed radiomic models based on CT image features and clinical parameters to distinguish between $EGFR^-$ and $EGFR^+$, and $KRAS^+$ and KRAS. Liu *et al*. [14] also, evaluated the capability of CT image features to predict EGFR mutation status in 298 surgically-resected peripheral lung adenocarcinomas in an Asian cohort of patients and build a high performance predictive model by using multiple logistic regression algorithm. Zhang *et al*. [15] also developed a radiogenomic model based on CT image features to predict EGFR and KRAS mutations in lung adenocarcinoma patients.

In the present work, we aimed to develop predictive models using PET, CT, and PET/CT image radiomic features with different machine learning approaches for optimal prediction of KRAS and EGFR mutation status.

## II. METHODS

### A. Image Data-Set

We considered 211 NSCLS cancer patients, each including diagnostic CT (CTD), low dose CT (CTA) and PET, as well as mutation status for KRAS and EGFR. After applying including/excluding criteria, 186 patients manually segmented on PET images, 175 patients manually segmented on CTA & CTD.

### B. Image Pre-Processing

Pre-processing include Laplacian of Gaussian (LOG), wavelet decomposition (WAV) and discretization to 64 bin (BIN64) were applied alone or in combinations on images. For LOG filter, different sigma value used to extract fine, medium and coarse (sigma value 0.5 to 5 with 0.5 steps) features.

### C. Feature Extraction

After applying pre-processing filters, image dataset was prepared and several radiomic features from different feature sets were extracted. The main feature sets were including; first order statistics and SUV based (19 features), Shape-based (16 features), gray level co-occurrence matrix (GLCM; 23 features), gray level run length matrix (GLRLM; 16 features), gray level size zone matrix (GLSZM; 16 features), neighboring gray tone difference matrix (NGTDM; 5 features), and gray level dependence matrix (GLDM; 14 features).

### D. Machine Learning

In this study, we used different algorithms for data splitting, feature selection and classification. For feature selection, we used Select K Best (KB), Select K Best & Mutual Info Regression (KB-MIR), Select from Model (SM) and Variance Threshold & Select from Model (VT-SM).

For data splitting 10-fold cross validation were used. In the present work, we trained classifiers with two different methods; balance and un-balance. In balance method, we have equal data from each class, where as in un-balance, there is no equal data and data are selected randomly.

For classification, we used random forest (RF), Bagging (BG), Logistic Regression (LREG), Naïve Bayesian (NB) and Support Vector Machine (SVM). To develop predictive radiomic models, patients with mutation in KRAS and EGFR were considered as class with label (+) and no mutations were considered with label (-).

### E. Statistical Analysis

All analysis including feature selection and classification were performed by in house developed python codes. The prediction ability of model was checked using the area under the receiver operator characteristic (ROC) curve (AUC).

## III. RESULTS

Our results for EGFR mutation status prediction based on feature selection methods and image sets are depicted in Fig. 1-a (mean AUC). It can be deduced from Fig. 1-a that feature selection performance had a small range from 0.53 to 0.6, and the combination of SM+PET-WAV had the highest performance (AUC: 0.60).

Fig. 1-b shows our results regarding EGFR mutation status prediction based on classifier and image sets. According to these results, classification performance has a range from 0.50 to 0.67 and combination of NB+CTD-BIN-LOG-WAV had the highest performance (AUC: 0.67).

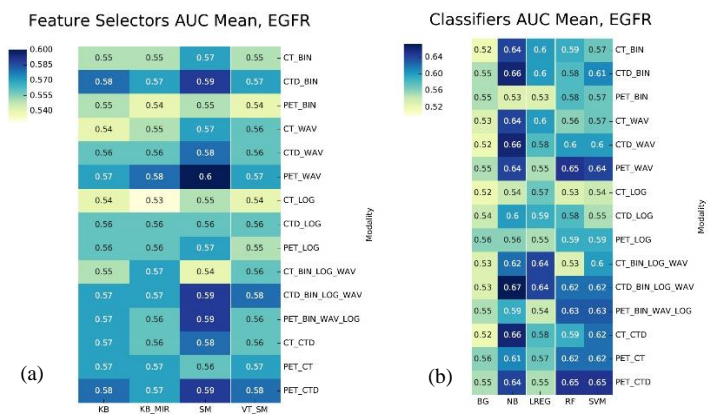

**Fig. 1.** Prediction of EGFR mutation status (in term of AUC), Rows; image sets Columns; a) feature selection b) Classifier

Fig. 2-a, show KRAS mutation status prediction results based on feature selection methods and image sets. In these results, feature selection performance has a small range from 0.52 to 0.56, and the combination of SM+PET-CT had the highest performance (AUC: 56).

KRAS mutation status prediction results based on feature classification and image sets is presented in Fig. 2-b (mean AUC). Here, the classification performance has a range from 0.50 to 0.62 and combination of SVM+PET-CT had the highest performance (AUC: 0.62).

clinical PET features. We further observed that the combination method CTD-BIN+B-KB+LREG (AUC: 0.75±0.10) had the highest predictive performance, followed by PET-CTD+B-VT-KB+MIR-SVM (AUC: 0.74±13), PET-CTD+B-KB+MIR-SVM (AUC: 0.74±13) lead to best performance in EGFR. Also, for KRAS CTD-BIN-LOG-WAV+B-VT-KB+MIR-LREG (AUC: 0.75±0.08) and CTD-BIN-LOG-WAV+B-VT+LREG (AUC: 0.75±0.07) had the best performance.

Our univariate analysis on conventional PET models (ROC curves in Fig. 3 for mutation status prediction showed that $SUV_{peak}$ (AUC=0.69) and MTV (AUC: 0.55) had the highest performance for EGFR and KRAS respectively.

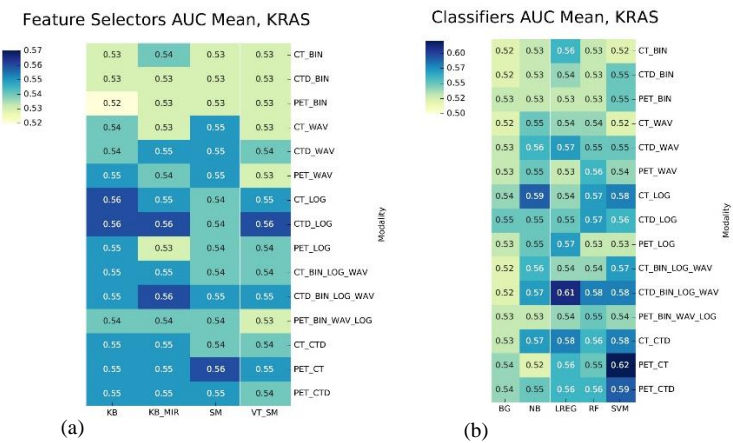

Fig. 2. Prediction of KRAS mutation status (in term of AUC), Rows; image sets Columns; a) feature selection b) classification

Fig. 3 depicts the ROC curves of the best predictive models in both machine learning and conventional

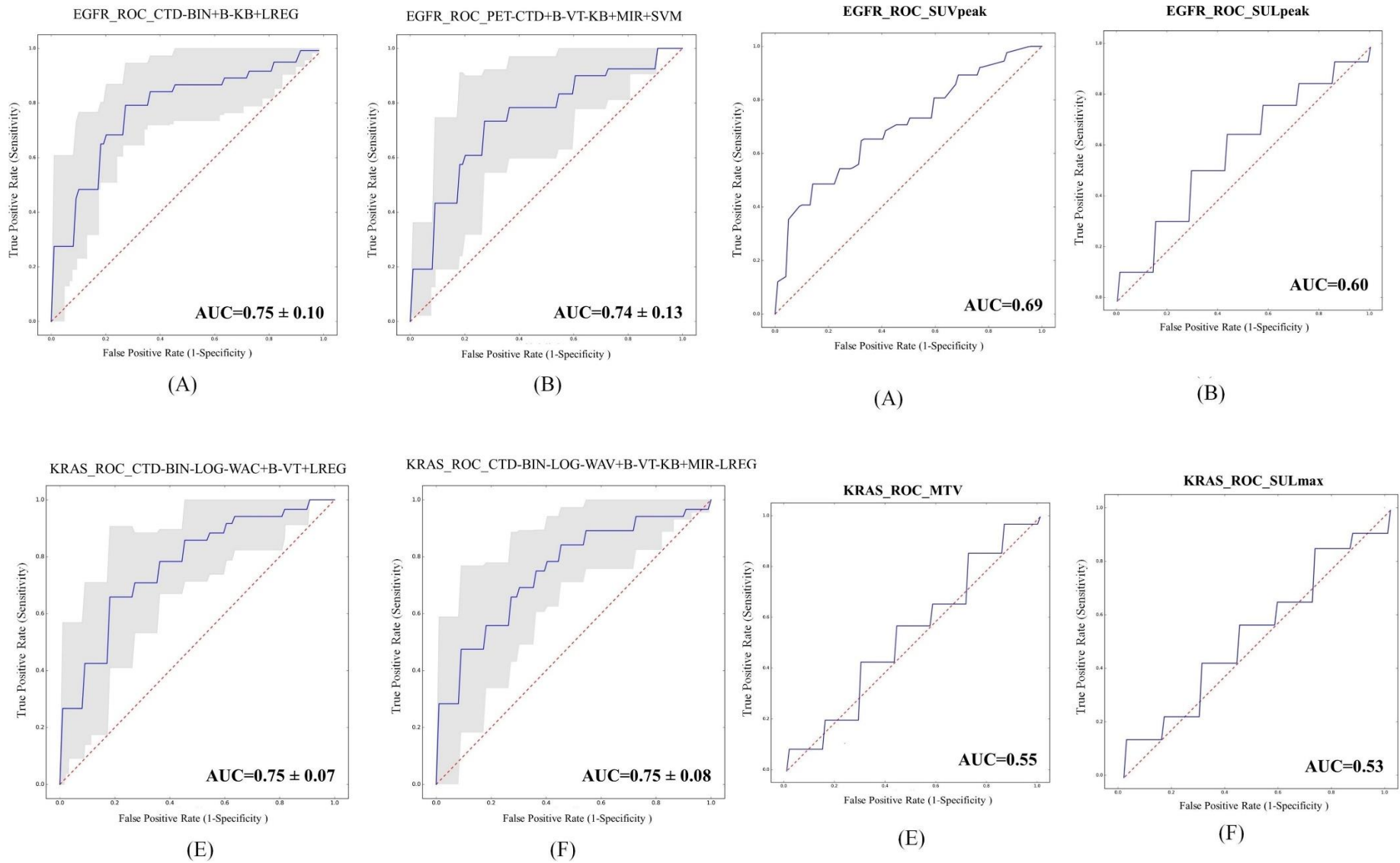

**Fig. 3.** ROC curve of prediction in machine learning algorithm (A, B) and conventional PET features (C, D), Top Row; EGFR, Bottom Row; KRAS, Error bars (shading region); 10-fold cross validation in machine learning algorithm

## IV. DISCUSSION

Predictive radiomic modeling is an active area of research in personalized medicine. In recent years a wide range of retrospective/prospective researches have been made on radiomics correlating phonotypical features with genomic status [1].

In the present study we developed radiomic signatures predictive of mutational status in NSCLC patients that showed good predictive performance in some imaging modalities. For EGFR and KRAS mutation status prediction a cross combination of imaging/ validation/ feature selection/ Classification methods resulted in a wide range of performance (AUC: 0.5~0.75).

In the present study, we developed sequencer radiomic models based on different image sets, pre-processing, feature selection, and classifier. This sequencing approach will result in a wide range of predictive models with different performances. In this situation, care should be taken into account for selecting the best models.

Some previous radiomics studies have tested different machine learning algorithms for introducing the best methods for feature selection and classifier[16]. Parmer *et al.* [17] by comparing fourteen feature selection and twelve classification methods in terms of their performance and stability found that WLCX (feature selection) and random forest (classification) had the highest prognostic performance with high stability against data perturbation. Zhang *et al.* [18], also, build a 54 cross combination machine learning algorithms including six feature selection and nine classification methods for survival prediction of advanced nasopharyngeal carcinoma. They found that the combination methods Random Forest (RF) and RF had the highest prognostic performance, followed by RF+Adaptive Boosting and Sure Independence Screening (SIS) + Linear Support Vector Machines.

In our results, we compared the conventional PET parameters (MTV, SUV and SUL) with radiomic features. We showed that radiomic models are more predictive than conventional clinical PET parameters. In our radiogenomic study, we found that also, the combination of PET and CT image features could better predict gene mutation status rather than parameters including SUV or MTV.

This study is showed, EGFR and KRAS mutation prediction using PET, CT and their combinations and be considered as a feasible and prospective approach for clinical decision making. In comparison with previous studies, our results seem to be consistent and comparable.

## V. Conclusion

In conclusion, the results presented here show the ability of PET, CT and PET/CT radiomic features to predict mutation status in cancer patients. This study incorporated PET images in lung cancer patients to find a link between PET/CT quantitation and genomics. The results indicate the potential of radiogenomics analysis to predict mutation statues in NSCLC PET/CT images in non-invasive manner.

# VI. REFERENCES

[1] K. Pinker, F. Shitano, E. Sala, R. K. Do, R. J. Young, A. G. Wibmer, H. Hricak, E. J. Sutton, and E. A. Morris, "Background, current role, and potential applications of radiogenomics," *Journal of Magnetic Resonance Imaging,* vol. 47, no. 3, pp. 604-620, 2018.

[2] P. J. Roberts, T. E. Stinchcombe, C. J. Der, and M. A. Socinski, "Personalized medicine in non–small-cell lung cancer: is KRAS a useful marker in selecting patients for epidermal growth factor receptor–targeted therapy?," *Journal of clinical oncology,* vol. 28, no. 31, pp. 4769-4777, 2010.

[3] H. Abdollahi, I. Shiri, and M. Heydari, "Medical Imaging Technologists in Radiomics Era: An Alice in Wonderland Problem," *Iranian journal of public health,* vol. 48, no. 1, pp. 184, 2019.

[4] H. Abdollahi, S. R. Mahdavi, I. Shiri, B. Mofid, M. Bakhshandeh, and K. Rahmani, "Magnetic resonance imaging radiomic feature analysis of radiation-induced femoral head changes in prostate cancer radiotherapy," *Journal of cancer research and therapeutics,* vol. 15, no. 8, pp. 11, 2019.

[5] R. J. Gillies, P. E. Kinahan, and H. Hricak, "Radiomics: images are more than pictures, they are data," *Radiology,* vol. 278, no. 2, pp. 563-577, 2015.

[6] I. Shiri, A. Rahmim, P. Ghaffarian, P. Geramifar, H. Abdollahi, and A. Bitarafan-Rajabi, "The impact of image reconstruction settings on 18F-FDG PET radiomic features: multi-scanner phantom and patient studies," *European radiology,* vol. 27, no. 11, pp. 4498-4509, 2017.

[7] I. Shiri, H. Abdollahi, S. Shaysteh, and S. R. Mahdavi, "Test-retest reproducibility and robustness analysis of recurrent glioblastoma MRI radiomics texture features," *Iranian Journal of Radiology*, no. 5, 2017.

[8] A. Zwanenburg, S. Leger, M. Vallières, and S. Löck, "Image biomarker standardisation initiative," *arXiv preprint arXiv:1612.07003*, 2016.

[9] L. Lu, W. Lv, J. Jiang, J. Ma, Q. Feng, A. Rahmim, and W. Chen, "Robustness of Radiomic Features in [11 C] Choline and [18 F] FDG PET/CT Imaging of Nasopharyngeal Carcinoma: Impact of Segmentation and Discretization," *Molecular Imaging and Biology,* vol. 18, no. 6, pp. 935-945, 2016.

[10] H. Abdollahi, S. Mostafaei, S. Cheraghi, I. Shiri, S. R. Mahdavi, and A. Kazemnejad, "Cochlea CT radiomics predicts chemoradiotherapy induced sensorineural hearing loss in head and neck cancer patients: a machine learning and multi-variable modelling study," *Physica Medica,* vol. 45, pp. 192-197, 2018.

[11] J. Beik, M. B. Shiran, Z. Abed, I. Shiri, A. Ghadimi-Daresajini, F. Farkhondeh, H. Ghaznavi, and A. Shakeri-Zadeh, "Gold nanoparticle-induced sonosensitization enhances the antitumor activity of ultrasound in colon tumor-bearing mice," *Medical physics,* vol. 45, no. 9, pp. 4306-4314, 2018.

[12] C. Parmar, P. Grossmann, D. Rietveld, M. M. Rietbergen, P. Lambin, and H. J. Aerts, "Radiomic machine-learning classifiers for prognostic biomarkers of head and neck cancer," *Frontiers in oncology,* vol. 5, pp. 272, 2015.

[13] E. R. Velazquez, C. Parmar, Y. Liu, T. P. Coroller, G. Cruz, O. Stringfield, Z. Ye, M. Makrigiorgos, F. Fennessy, and R. H. Mak, "Somatic mutations drive distinct imaging phenotypes in lung cancer," *Cancer research,* vol. 77, no. 14, pp. 3922-3930, 2017.

[14] Y. Liu, J. Kim, Y. Balagurunathan, Q. Li, A. L. Garcia, O. Stringfield, Z. Ye, and R. J. Gillies, "Radiomic features are associated with EGFR mutation status in lung adenocarcinomas," *Clinical lung cancer,* vol. 17, no. 5, pp. 441-448. e6, 2016.

[15] L. Zhang, B. Chen, X. Liu, J. Song, M. Fang, C. Hu, D. Dong, W. Li, and J. Tian, "Quantitative Biomarkers for Prediction of Epidermal Growth Factor Receptor Mutation in Non-Small Cell Lung Cancer," *Translational oncology,* vol. 11, no. 1, pp. 94-101, 2018.

[16] H. Abdollahi, B. Mofid, I. Shiri, A. Razzaghdoust, A. Saadipoor, A. Mahdavi, H. M. Galandooz, and S. R. Mahdavi, "Machine learning-based radiomic models to predict intensity-modulated radiation therapy response, Gleason score and stage in prostate cancer," *La radiologia medica*, pp. 1-13, 2019.

[17] C. Parmar, P. Grossmann, J. Bussink, P. Lambin, and H. J. Aerts, "Machine learning methods for quantitative radiomic biomarkers," *Scientific reports,* vol. 5, pp. 13087, 2015.

[18] B. Zhang, X. He, F. Ouyang, D. Gu, Y. Dong, L. Zhang, X. Mo, W. Huang, J. Tian, and S. Zhang, "Radiomic machine-learning classifiers for prognostic biomarkers of advanced nasopharyngeal carcinoma," *Cancer letters,* vol. 403, pp. 21-27, 2017.